\documentclass[fleqn,5p]{elsarticle}

\journal{Sustainable Energy, Grids and Networks}

\usepackage{multicol}
\usepackage{amsmath}
\usepackage{amssymb}
\usepackage{mathtools}
\usepackage{cuted}
\usepackage{graphicx}
\usepackage{flafter}
\usepackage{placeins}
\usepackage{IEEEtrantools}
\usepackage{textcomp}
\usepackage{hyperref}
\usepackage{url}       
\usepackage{color}
\usepackage{tabularx}
\usepackage{booktabs}
\usepackage{multirow}
\usepackage{pgfplots}
\pgfplotsset{compat=1.18}
\usepackage{xcolor}
\usepackage[table]{xcolor}
\usepackage{caption}

\usepackage{booktabs}
\usepackage{pifont}
\newcommand{\cmark}{\ding{51}} 
\newcommand{\xmark}{\ding{55}} 
\newcommand{\tmark}{\texttildelow} 

\usepackage{xcolor}

\hypersetup{urlcolor=cyan}

\begin{document}

\setlength{\tabcolsep}{6pt}
\setlength{\mathindent}{0pt}

\begin{frontmatter}

\title{Least-Cost Overvoltage Control in PV-Rich Distribution Networks\\via Unbalanced Optimal Power Flow}

\author[UPM]{A. Espinosa Del Pozo}
\ead{andrea.espinosa.delpozo@upm.es}
\author[UPM]{A. Hernández}
\ead{araceli.hernandez@upm.es}
\author[UPM]{L. Badesa}
\ead{luis.badesa@upm.es}
\address[UPM]{Universidad Politécnica de Madrid, Spain}

\begin{abstract}
The increasing penetration of photovoltaic (PV) generation in low-voltage distribution networks presents operational challenges, with overvoltages being among the most critical. This study introduces a tool based on Unbalanced Optimal Power Flow (UBOPF) to assess cost-effective local inverter control strategies specifically aimed at mitigating overvoltage issues. Two approaches are examined:~dynamic active power curtailment and combined active and reactive power control. These strategies are tested on a residential low-voltage network with high PV penetration, where the UBOPF model with voltage-magnitude constraints was implemented in \texttt{Julia} using the \texttt{JuMP} optimization package. The results demonstrate that both methods are effective in maintaining voltage levels within regulatory limits, with the latter leading to lower PV curtailment. The analysis highlights the need to consider these control actions as ancillary services to the grid, which should be properly compensated given their effect on generator revenues. 

\end{abstract}

\begin{keyword}
Unbalanced Optimal Power Flow (UBOPF) \sep Overvoltage Mitigation \sep Low-Voltage Distribution Networks \sep Inverter-Based Control Strategies \sep Ancillary Services 
\end{keyword}

\end{frontmatter}

\section{Introduction} \label{Intro}
The increasing integration of distributed photovoltaic (PV) generation in low-voltage (LV) residential distribution networks poses significant challenges to voltage regulation and network stability \cite{Li}. Unlike medium-voltage grids, LV systems are characterized by high R/X ratios and predominantly radial topologies, which make them particularly sensitive to voltage rise phenomena. Overvoltage conditions typically arise when local PV generation exceeds household demand, especially during midday hours, and are exacerbated by the limited voltage control capabilities in these networks.

Several technical approaches have been proposed in the literature to mitigate overvoltage events in LV systems, including network reinforcement, installation of on-load tap changers or deployment of local energy storage systems \cite{Hashemi, Haque}. While effective, these measures often involve high capital expenditure and long deployment times, making them less viable for widespread residential application. In contrast, inverter-based voltage control strategies—such as active power curtailment and reactive power control—have received increasing attention due to their low cost, fast response and ease of implementation \cite{Hashemi, Ren}. These strategies can be deployed either locally or under centralized coordination and the choice between both approaches remains an active topic of research. Some studies highlight the simplicity and scalability of decentralized control, while others emphasize the benefits of coordinated schemes in terms of global efficiency and system observability \cite{Boli, Garcia, Gui}. In addition, rapidly changing irradiance conditions may lead to voltage fluctuations that cannot be addressed by conventional utility equipment, highlighting the potential role of fast-responding, var-capable inverters in supporting voltage regulation under dynamic operating conditions \cite{Turitsyn}.

To evaluate and design such inverter-based control strategies, Optimal Power Flow (OPF) formulations have been widely adopted in the context of distribution networks. Although multiple approaches can be used to address voltage regulation problems, an OPF-based formulation is adopted in this work in order to enable the mitigation of overvoltage conditions in a cost-effective manner while maximizing the integration of renewable generation. The OPF framework offers a principled way to coordinate inverter operation under network constraints, ensuring efficient and scalable use of distributed resources. Several studies have applied OPF models to the analysis and planning of voltage regulation in distribution networks, often focusing on medium-voltage feeders and assuming balanced network conditions \cite{Ceylan, Meirinhos}. While these approaches provide valuable insights, their underlying assumptions limit their accuracy when applied to LV residential systems.

In LV networks, residential loads and PV installations are typically single-phase and unevenly distributed across the three phases, giving rise to significant phase-to-phase voltage differences. These unbalanced conditions cannot be captured by conventional balanced OPF formulations~\cite{Geth}. Accurately assessing inverter-based voltage control strategies in this context requires modelling tools that represent the per-phase behaviour of both the network and distributed energy resources (DER). This need has led to the development of Unbalanced Optimal Power Flow (UBOPF) formulations, which extend classical OPF models to include phase-specific voltages and asymmetrical loading. A comprehensive review of modelling approaches for LV distribution systems and unbalanced optimal power flow formulations, including their mathematical foundations, solution techniques and main application areas, is provided in \cite{Ibrahim}. Similar mathematical descriptions of unbalanced LV networks can also be found in~\cite{Kryonidis}, which adopts a comparable three-phase representation to enable distributed reactive power control. In parallel, recent advances in three-phase inverter control have also focused on mitigating current unbalance and improving voltage quality in LV grids, as demonstrated by Lago et al.~\cite{Lago}. Building upon this body of work, subsequent research has applied unbalanced OPF-based approaches to feeders with high PV penetration, enabling more realistic per-phase analysis of voltage-support actions~\cite{Lin}.

Within the OPF/UBOPF literature, recent contributions have explored different modelling and coordination choices for inverter-based voltage support. These include convexified or linearized formulations coordinating inverter support (and, in some cases, utility devices) under voltage constraints~\cite{LiDisfani}, nonlinear three-phase AC-OPF formulations applied to PV-rich distribution systems under alternative operational objectives such as CVR~\cite{Gutierrez}, and OPF-based studies that benchmark centralized nonlinear optimization against sensitivity-based coordination~\cite{Ceylan}. Related OPF formulations are also commonly developed under balanced assumptions or medium-voltage settings~\cite{Meirinhos} and sensitivity-aware dispatch approaches have been proposed for smart inverters with multiple control modes~\cite{Almomani}.

Alongside OPF/UBOPF formulations, a second research stream focuses explicitly on inverter control strategies for voltage regulation in PV-rich networks without solving a full OPF problem. Distributed and local inverter controls have been proposed to mitigate voltage violations through reactive power support and, in some cases, active power curtailment, including inter-phase coordination schemes for unbalanced LV systems~\cite{Wang}, distributed Volt/Var control~\cite{Rushikesh}, zonal Volt/Var mechanisms for scalability under high PV penetration~\cite{Alrushoud} and enhanced local Var/Watt rules aligned with grid-code practice~\cite{Ghasemi}. Data-driven and hybrid inverter coordination strategies have also been explored to cope with uncertainty and fast PV dynamics~\cite{Vergara, Zhang}. Table~\ref{tab:comparison_expanded} summarizes representative contributions from both streams and contrasts them with the UBOPF framework proposed in this paper, highlighting the network modelling detail (balanced versus unbalanced three-phase), the use of nonlinear AC power flow equations (versus relaxations/approximations or rule-based control), the inverter actions considered ($P$ curtailment and/or $Q$ support) and the inclusion of an explicit economic remuneration layer —a combination that is not jointly addressed by the reviewed literature and constitutes the key gap targeted in this work.

\begin{table*}[t]
\centering
\caption{Expanded comparison of inverter-based voltage-management approaches for PV-rich distribution networks, with emphasis on how inverter control actions (active-power curtailment and reactive-power support) are modelled and coordinated.
Symbols: \cmark~explicitly modelled/considered; \xmark~not considered; \tmark~partially considered or based on approximations/relaxations.}
\label{tab:comparison_expanded}
\renewcommand{\arraystretch}{1.06}
\setlength{\tabcolsep}{4.0pt}
\scriptsize

\resizebox{\textwidth}{!}{%
\begin{tabular}{l l l c c c c c c l}
\toprule
\textbf{Reference} &
\textbf{Coord.} &
\textbf{Core method} &
\textbf{3$\phi$ unbal.} &
\textbf{Nonlinear AC} &
\textbf{Volt. constr.} &
\textbf{Inv. P} &
\textbf{Inv. Q} &
\textbf{Econ. layer} &
\textbf{Primary aim} \\
\midrule

\multicolumn{10}{l}{\textit{OPF/UBOPF-based}}\\
\midrule

Ceylan (2021)~\cite{Ceylan} &
Hybrid & NLP OPF vs sensitivity-based &
\xmark & \cmark & \cmark & \cmark & \cmark & \xmark &
Centralized vs local control (balanced feeder) \\

Meirinhos (2017)~\cite{Meirinhos} &
Centralized & Multi-temporal OPF (MV) &
\xmark & \cmark & \cmark & \cmark & \cmark & \tmark &
Day-ahead voltage control (MV, balanced) \\

Lin (2022)~\cite{Lin} &
Centralized & VSC-OPF via SDP (stability) &
\cmark & \tmark & \cmark & \tmark & \tmark & \xmark &
Voltage stability margin \\

Li (2019)~\cite{LiDisfani} &
Centralized & Convex/linearized coordination &
\cmark & \xmark & \cmark & \xmark & \cmark & \xmark &
OLTC + inverter Q coordination \\

Guti\'errez-Lagos (2019)~\cite{Gutierrez} &
Centralized & 3$\phi$ AC-OPF (CVR) &
\cmark & \cmark & \cmark & \xmark & \xmark & \tmark &
CVR / energy reduction (MV--LV) \\

Almomani (2025)~\cite{Almomani} &
Centralized & Linear OPF / sensitivity-aware &
\cmark & \xmark & \cmark & \xmark & \cmark & \xmark &
Reactive dispatch / setpoints \\
\midrule

\multicolumn{10}{l}{\textit{Control/coordination-oriented (no full AC-OPF/UBOPF core)}}\\
\midrule

Wang (2020)~\cite{Wang} &
Distributed & Inter-phase coordination (Volt/Var) &
\cmark & \xmark & \cmark & \xmark & \cmark & \xmark &
Inter-phase voltage control \\

Babu--Khatod (2024)~\cite{Rushikesh} &
Distributed & Distributed Volt/Var control &
\cmark & \xmark & \cmark & \xmark & \cmark & \xmark &
Voltage-violation mitigation \\

Alrushoud (2021)~\cite{Alrushoud} &
Distributed & Zonal Volt/VAR (clustering) &
\cmark & \xmark & \cmark & \xmark & \cmark & \xmark &
Fast scalable coordination \\

Ghasemi (2025)~\cite{Ghasemi} &
Local & Enhanced Var/Watt rules &
\tmark & \xmark & \tmark & \cmark & \cmark & \xmark &
Local OV mitigation (rules) \\

Vergara (2022)~\cite{Vergara} &
Hybrid & RL + NLP benchmark &
\cmark & \tmark & \cmark & \cmark & \xmark & \xmark &
Data-driven OV mitigation \\

Zhang (2025)~\cite{Zhang} &
Hybrid & Sensitivity + event-triggered control &
\xmark & \xmark & \cmark & \tmark & \tmark & \xmark &
Multi-time-scale voltage regulation (balanced) \\

Alabri (2021)~\cite{Alabri} &
Supervisory & Q-support (voltage + unbalance) &
\cmark & \xmark & \cmark & \xmark & \cmark & \xmark &
Voltage quality / VUF \\

Long (2022)~\cite{Long} &
Distributed & Coalition formation coordination &
\xmark & \xmark & \cmark & \xmark & \tmark & \xmark &
Coordinated voltage regulation (balanced) \\

Cheng (2020)~\cite{Cheng} &
Distributed & Distributed coordination (MAS/droop) &
\xmark & \xmark & \cmark & \xmark & \tmark & \xmark &
Coordination of many PVs (balanced) \\

\midrule

\textbf{This work} &
\textbf{Centralized} & \textbf{Nonlinear UBOPF} &
\textbf{\cmark} & \textbf{\cmark} & \textbf{\cmark} &
\textbf{\cmark} & \textbf{\cmark} & \textbf{\cmark} &
\textbf{Hosting capacity + remunerated service} \\
\bottomrule
\end{tabular}%
}
\end{table*}

Building upon these insights, this work introduces a comprehensive modelling and analysis framework based on the UBOPF formulation, specifically designed to capture the unbalanced operating conditions of LV distribution networks. The model includes a detailed three-phase representation of network elements, phase-specific voltage magnitude constraints and inverter capability limits. It enables inverter-based voltage control actions---namely dynamic active power curtailment and coordinated $P$--$Q$ control---to be represented within a single optimization framework. It should be noted that the proposed UBOPF model is intended for analysis purposes and assumes full knowledge of the network electrical parameters, centralized control, and idealized operating conditions. Its primary objective is to identify technically optimal inverter coordination strategies that maximize renewable integration while maintaining voltage levels within acceptable limits, rather than to define a real-time operational control scheme.

Consistent with Table~\ref{tab:comparison_expanded}, while existing studies address complementary aspects of inverter-based voltage regulation (e.g., local and distributed control laws, convexified OPF coordination, or OPF-based operation under different objectives), to the best of the authors' knowledge the literature lacks a unified framework that simultaneously (i) adopts a full nonlinear unbalanced three-phase OPF model tailored to LV residential feeders, (ii) coordinates both active power curtailment and reactive power support under explicit phase-specific voltage constraints, and (iii) incorporates an explicit economic layer that frames inverter-based voltage support as a remunerated ancillary service provided by PV owners and procured by the Distribution System Operator (DSO).

A key distinctive feature of the proposed approach is that these control strategies are not addressed merely as technical solutions to local voltage issues, but rather as economically motivated mechanisms. They are conceived as ancillary services provided by distributed PV generators and remunerated by the Distribution System Operator. Framing local voltage regulation in this way enables a shift toward market-based approaches that incentivize DER participation, supporting network reliability in high-PV environments.

The complete framework is implemented in the \texttt{Julia} programming language using the \texttt{JuMP} optimization environment and the non-linear solver \texttt{Ipopt}. Two inverter-based control strategies are analyzed: (i) dynamic active power curtailment, and (ii) combined active and reactive power control. Both strategies are embedded within the UBOPF formulation and applied to a representative 18-node LV residential feeder with high PV penetration. The model is used to evaluate the effectiveness of each strategy under varying operating conditions, with particular attention to the influence of phase imbalance, load power factor and the ratio between local demand and PV generation on voltage regulation performance.

The main contributions of this work are as follows:
\begin{itemize}
    \item For the first time, local inverter-based control strategies are embedded within an unbalanced optimal power flow (UBOPF) formulation as explicit constraints under phase-specific voltage magnitude limits, enabling the development of regulatory policies and incentive schemes for distributed PV integration in low-voltage networks.
   
    \item The effectiveness of different inverter-based control strategies is evaluated using the UBOPF, demonstrating on a representative base scenario that combining active and reactive power control reduces PV curtailment by up to 40\% compared to active power limitation alone, while effectively maintaining voltage levels within statutory limits.
    
    \item A novel quantitative analysis of the adverse effects of capacitive load behaviour and phase imbalance on voltage rise and PV hosting capacity in residential low-voltage networks is conducted. 
\end{itemize}

The remainder of this paper is structured as follows. Section~\ref{sec:overvoltages} provides background on overvoltage phenomena in low-voltage networks with high photovoltaic penetration and introduces the two inverter-based control strategies considered in this study. Section~\ref{sec:ubopf} outlines the formulation of the  UBOPF model, detailing the network representation, control constraints and optimization problem. Section~\ref{sec:case_studies} presents the application of the UBOPF framework to a test network, assessing the performance of the proposed control strategies and quantifying the influence of critical factors such as phase imbalance, load power factor, and local generation-to-demand ratio. Finally, Section~\ref{sec:conclusion} summarizes the main findings and discusses directions for future research.

\section{Overvoltage Phenomena in Low-Voltage Networks and Inverter-Based Control Strategies} \label{sec:overvoltages}

This section introduces the overvoltage phenomenon in low-voltage distribution systems with high photovoltaic penetration and then presents two inverter-based control strategies designed to mitigate it: dynamic active power reduction and combined active and reactive power control. The section also outlines key implementation aspects and regulatory considerations related to providing voltage support as an ancillary service.

\subsection{Overvoltage Phenomena in LV Distribution Networks}

In conventional low-voltage (LV) distribution networks, voltage magnitude progressively decreases from the transformer secondary to the end-users due to the resistive nature of the conductors. However, with the widespread installation of rooftop PV systems, this trend can be reversed during periods of high irradiance and low local demand, when surplus active power is injected into the grid. Under these conditions, reverse power flow raises the voltage at the point of common coupling (PCC) and may propagate upstream, potentially violating operational voltage limits and triggering automatic disconnection of PV inverters.

This phenomenon can be described using a Thevenin equivalent representation of the upstream grid, where the PCC voltage depends on the net injected active and reactive power and the equivalent network impedance. Following \cite{Hashemi}, the voltage rise at the PCC can be approximated by 

\begin{equation} \Delta V \approx \frac{P \cdot R_{\text{th}} + Q \cdot X_{\text{th}}}{U_{\text{th}}} \label{eq:volt_rise} \end{equation} where \( \Delta V = |\underline{U}_{\text{PCC}} - \underline{U}_{\text{th}}| \) is the voltage rise caused by the net injection of active (\(P\)) and reactive (\(Q\)) power into the grid, \( R_{\text{th}} \) and \( X_{\text{th}} \) are the equivalent resistance and reactance between the PCC and the transformer, and \( U_{\text{th}} \) is the Thevenin voltage.

This linear approximation clearly highlights the distinct influence of active and reactive power on voltage regulation. In LV networks, where \(R/X \gg 1\), the resistive component dominates, so variations in active power (\(P\)) exert a much stronger impact on voltage magnitude than reactive power (\(Q\)). Although the effect of reactive power is comparatively smaller, it can be effectively exploited by inverters to provide dynamic voltage support. 

Regulatory standards, such as UNE-EN~50160 \cite{UNE}, define permissible voltage ranges (typically \(\pm10\%\) of nominal), but many DSOs apply tighter operational limits (e.g., \(\pm3\%\)) to avoid inverter disconnection events \cite{Hashemi, Stetz}. Simulations on real LV networks have shown that voltage rise can reach 1--3\% per 30~kVA of distributed generation, depending on system impedance and the location of the PV systems \cite{Stetz}. This represents a practical limit to the hosting capacity of the network and can prevent the connection of additional renewable sources unless mitigation strategies are implemented.

\subsection{Inverter-Based Strategies for Overvoltage Mitigation}

Most PV inverters in self-consumption systems operate to maximize active power injection at unity power factor, as only energy is typically remunerated. While this maximizes returns for prosumers, it can exacerbate voltage rise under high generation and low demand conditions. These situations do not indicate a fault but reflect the cumulative effect of local generation exceeding the network’s absorption capability.

This study addresses two control strategies that can be implemented locally in PV inverters to mitigate overvoltages in low-voltage networks, without compromising the overarching goal of maximizing renewable energy integration. Both strategies are included within the UBOPF framework, where voltage constraints are explicitly enforced and the objective function is designed to maximize the total renewable generation injected into the system.

\paragraph{Dynamic Active Power Curtailment}

This approach allows inverters to inject only active power at unity power factor but introduces a dynamic upper bound on the active power output, which depends on the proximity to the voltage limit. The inverter reduces its active power injection as needed to keep local voltages within acceptable bounds. This strategy does not require reactive power control capability or modification of inverter logic and serves as a minimum intervention strategy. It also enables the quantification of PV curtailment due to voltage constraints.

Under unity power factor operation, the inverter effectively operates along the vertical axis of the $P$-$Q$ plane, injecting only active power. Its operation is constrained by the maximum available apparent power \( S_{\text{PV,max}} \), which depends on instantaneous solar irradiance. This instantaneous limit should not be confused with the inverter’s nominal rating, as the full apparent power capacity is not always available; rather, irradiance determines the actual amount of power that can be delivered at any given moment. Consequently, the inverter adjusts its active power output within this irradiance-dependent limit, reducing the injected active power when necessary to maintain acceptable voltage levels while continuing to operate at unity power factor.

\paragraph{Combined Active and Reactive Power Control}

In this strategy, inverters operate within their full apparent power capability. They can reduce active power output and absorb reactive power in order to mitigate voltage rise. As shown in the voltage rise expression~\eqref{eq:volt_rise}, reactive power absorption contributes to reducing the voltage magnitude. According to the sign convention used in that expression, negative reactive power represents power flowing from the grid into the inverter (i.e., absorption), while positive values correspond to injection into the grid.

The UBOPF model treats both \( P_{\text{PV}} \) and \( Q_{\text{PV}} \) as decision variables, constrained by the inverter’s $P$–$Q$ capability plane defined by the apparent power limit:
\begin{equation}
P_{\text{PV}}^2 + Q_{\text{PV}}^2 \leq S_{\text{PV,max}}^2,
\end{equation}
which ensures that the operating point of the inverter always remains within its instantaneous available apparent power capacity, determined by the solar irradiance at that moment. This added flexibility often allows for higher active power injection compared to the previous strategy, by dedicating part of the inverter’s capacity to voltage support through reactive power control. As a consequence, the inverter does not operate at a fixed power factor; instead, its power factor varies dynamically depending on the balance between active power generation and the reactive support required to maintain voltage levels within limits.

Under this combined control approach, the inverter can move to any internal point of the capability region defined by the semicircle \(P_{\text{PV}}^2 + Q_{\text{PV}}^2 \leq S_{\text{PV,max}}^2\). This region represents all feasible combinations of active and reactive power that respect the apparent power limit imposed by the available irradiance. By allocating part of its capacity to reactive power absorption, the inverter gains the ability to reduce \(\Delta V\) more effectively, particularly when operating near its active power limit. As a result, the inverter can simultaneously inject active power and provide voltage support, selecting the most suitable combination of \(P_{\text{PV}}(t)\) and \(Q_{\text{PV}}(t)\) to maintain voltage magnitudes within permissible bounds.

\begin{figure}[h!]
    \centering
    \includegraphics[width=0.32\textwidth]{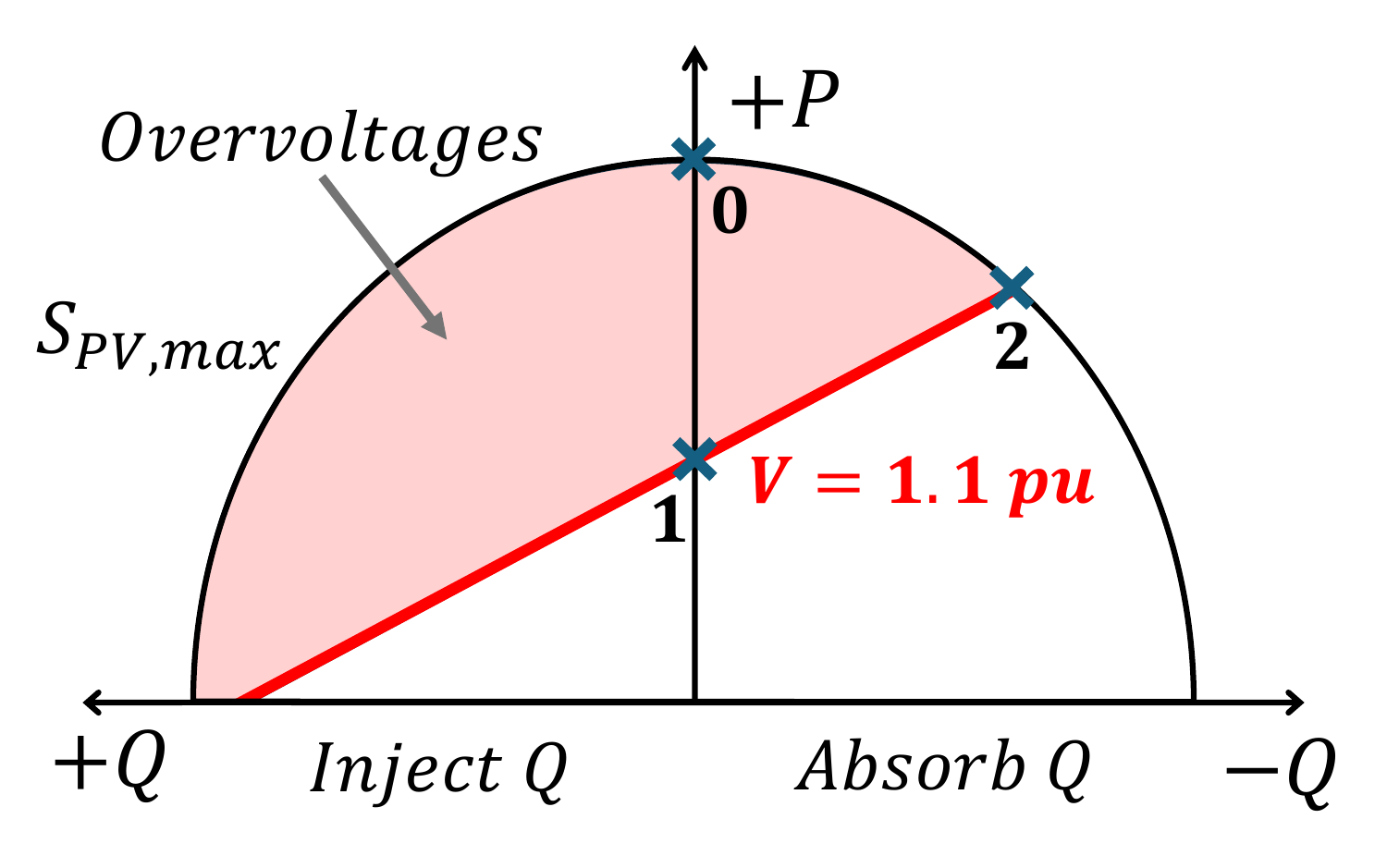}
    \caption{Operating points under different inverter control strategies with a local voltage constraint of \( V = 1.1\,\mathrm{pu} \) shown as a straight line in the \(P\)--\(Q\) plane. Point~0: full active power injection without control, leading to overvoltage. Point~1: active power curtailment at unity power factor. Point~2: combined active and reactive power control to maximize active power injection while remaining within voltage limits.}
    \label{fig11}
\end{figure}

To further clarify the operating principles of the proposed inverter-based control strategies, a graphical example is presented in Figure~\ref{fig11}, illustrating their effect on the inverter’s operating point. 
In this example, a voltage constraint of \( V = 1.1\,\mathrm{pu} \) is enforced specifically on the voltage magnitude at the inverter terminal, leading to a linear boundary in the \(P\)--\(Q\) plane \cite{Domenech}. Any operating point located above this line results in a voltage magnitude that exceeds the permissible limit.

Point~0 corresponds to the scenario without any control strategy. Here, the inverter injects the full available apparent power \( S_{\text{PV,max}} \) entirely as active power at unity power factor. While this operating point maximizes energy delivery, it leads to overvoltage due to the absence of voltage regulation. Under the dynamic active power curtailment strategy, the inverter maintains unity power factor but reduces its active power output to avoid violating the voltage constraint. This behaviour is represented by point~1, which lies on the boundary defined by the voltage limit. Although this strategy mitigates overvoltage, it does not exploit the inverter’s capability to provide reactive support, which could further reduce curtailment.

Point~2 illustrates the effect of the combined active and reactive power control strategy. In this case, the inverter absorbs reactive power (\( Q_{\text{PV}} < 0 \)), which lowers the local voltage and allows for greater active power injection without exceeding the voltage constraint. By actively managing both power components within the $P$-$Q$ plane, this strategy becomes more effective in reducing PV curtailment. This advantage will be quantitatively confirmed in the case studies presented in Section~\ref{sec:case_studies}.

\subsection{Local vs.~Centralized Control Implementation}

Both control strategies can be implemented using either local or centralized schemes. In local control, the inverter reacts autonomously to local voltage measurements, adjusting its output without external communication. This method is scalable and robust but can result in uncoordinated behaviour among devices. Centralized control, by contrast, involves the DSO issuing real-time setpoints to inverters based on global network conditions. This allows for optimized coordination but requires communication infrastructure and control protocols.

Hybrid approaches may also be adopted, where local control provides fast primary response and centralized control adjusts operating limits or setpoints periodically. Although this paper focuses on offline optimization, the control actions simulated in the UBOPF model can be interpreted under either control paradigm.

\subsection{Regulatory and Economic Considerations}

The adoption of inverter-based voltage control strategies, such as active power curtailment or reactive power injection, plays a key role in mitigating overvoltages in LV distribution networks with high PV penetration. From a technical perspective, these control actions enhance voltage stability and allow greater integration of distributed energy resources. However, from the viewpoint of prosumers, they imply a reduction in the energy self-consumed or exported to the grid, particularly in the case of active power curtailment, which directly affects their economic return.

Two main regulatory approaches have emerged to promote the use of such control functionalities. The first consists of enforcing specific behaviours through technical standards or grid codes. For example, several countries have made reactive power control capabilities mandatory for new PV installations, requiring inverters to operate under fixed power factor or voltage-reactive power Q(V)-curves. These are however rarely applicable to domestic PV assets. The second approach involves the economic valuation of these services, by recognizing them as ancillary services provided to the distribution system. In this context, DER contribute to maintaining voltage profiles within operational limits and potentially reduce the need for network reinforcements, which justifies the design of compensation mechanisms. The perspective adopted in this work aligns with this second approach, viewing inverter-based control not only as a technical solution but also as a service whose value to the grid should be recognized and remunerated.

While current regulatory frameworks are still evolving, the inclusion of such control actions in optimization-based tools like the Unbalanced Optimal Power Flow enables a quantitative assessment of their economic impact. Specifically, the UBOPF framework allows for the evaluation of how different control policies affect voltage regulation and the maximum amount of renewable energy that can be integrated without violating network constraints. Although economic incentives are not explicitly modelled in the UBOPF formulation, the results can inform cost-benefit analyses by identifying the trade-offs between grid stability and curtailed energy under various operational scenarios. This provides a valuable basis for designing remuneration schemes aligned with the actual value of inverter-based services to the grid.

\section{UBOPF Formulation} \label{sec:ubopf}

This section presents the formulation of the Unbalanced Optimal Power Flow problem for multi-phase distribution networks. The formulation is based on the nodal admittance model using complex quantities and explicitly accounts for the unbalanced and coupled nature of LV feeders. This formulation builds upon the modelling framework proposed in~\cite{Yang}, adapting it to the specific context and objectives of this work. Our code is freely available in~\cite{github_repo}.

This paper focuses exclusively on single-phase and three-phase wye-connected loads and generators, for which power generation and consumption can be modelled independently in each phase. Further investigation is needed to integrate delta-connected elements into this UBOPF formulation.

Throughout this section, different typographic conventions are adopted to clearly distinguish the mathematical nature of each quantity. Complex scalars are indicated by an underline, such as in the nodal voltage \(\underline{V}_i^\phi\); real-valued vectors and matrices are written in bold, e.g., \(\mathbf{P}_{G,i}\) or \(\mathbf{P}_{D,i}\); and complex vectors or matrices are expressed in bold and underlined, such as \(\underline{\mathbf{V}}\) and \(\underline{\mathbf{Y}}\).

\subsection{Nomenclature}

\subsubsection*{Indices and Sets}
\begin{tabularx}{\columnwidth}{l X}
$i,j$ & Node indices \\
$\phi$ & Phase index, $\phi \in \{a, b, c\}$ \\
$\mathcal{N}$ & Set of nodes \\
$\mathcal{L}$ & Set of electrical lines, defined as pairs of connected nodes $(i,j)$ \\
$\mathcal{G}_c$ & Set of nodes with conventional generation \\
$\mathcal{G}_{\text{pv}}$ & Set of nodes with PV generation \\
\end{tabularx}

\vspace{1ex}
\subsubsection*{Variables}
\begin{tabularx}{\columnwidth}{l X}
$\underline{\mathbf{V}}$ & Global complex voltage vector \\
$\underline{V}_i^{\phi}$ & Complex voltage at node $i$, phase $\phi$ \\
$P_{G,i}^{\text{conv},\phi}, Q_{G,i}^{\text{conv},\phi}$ & Active/reactive power from conventional generator at node $i$, phase $\phi$ \\
$P_{G,i}^{\text{pv},\phi}, Q_{G,i}^{\text{pv},\phi}$ & Active/reactive power from PV inverter at node $i$, phase $\phi$ \\
$\underline{I}_i^{\phi}$ & Complex injected current at node $i$, phase $\phi$ (linear expression) \\
$\underline{S}_{ij}^{\phi}$ & Complex power flow on line $(i,j)$, phase $\phi$ (linear expression) \\
\end{tabularx}

\vspace{1ex}
\subsubsection*{Parameters}
\begin{tabularx}{\columnwidth}{l X}
$P_{D,i}^{\phi}, Q_{D,i}^{\phi}$ & Active/reactive demand at node $i$, phase $\phi$ \\
$S_{G,i}^{\text{pv},\phi}$ & Apparent power limit of PV inverter \\
$P_{G,i}^{\text{conv},\phi,\min/\max}$ & Active power bounds of conventional generator \\
$Q_{G,i}^{\text{conv},\phi,\min/\max}$ & Reactive power bounds of conventional generator \\
$c_i^{\text{conv}}$ & Cost coefficient of conventional generator at node $i$ \\
$V_{\min}^{\phi}, V_{\max}^{\phi}$ & Voltage magnitude limits \\
${P}_{ij,max}^{\phi}$ & Maximum allowable active power flow in line $(i,j)$, phase $\phi$ \\
$\underline{\mathbf{Y}}$ & Global nodal admittance matrix \\
$\underline{\mathbf{Y}}_{ij}^{\text{br}}$ & Branch admittance matrix of line $(i,j)$ \\
$\underline{\mathbf{Z}}_{ij}$ & Series impedance matrix of line $(i,j)$ \\
$\underline{\mathbf{Y}}_{\text{sh},ij}$ & Shunt admittance matrix of line $(i,j)$ \\
\end{tabularx}

\subsection{Branch Admittance Matrix Construction}

\begin{figure}[h!]
    \centering
    \includegraphics[width=1\linewidth]{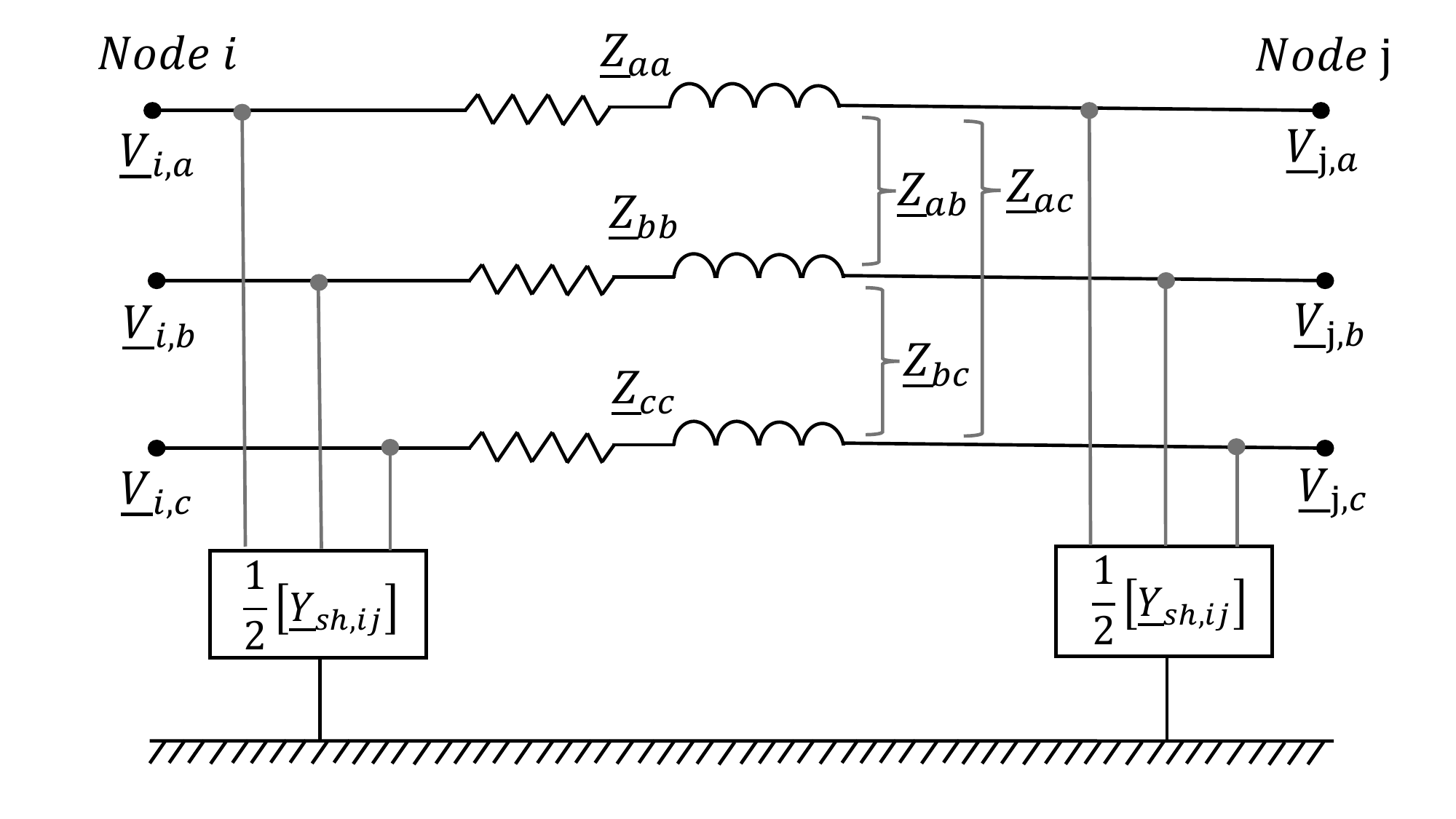}
    \caption{Three-phase $\pi$-model of a distribution line between nodes $i$ and $j$.}
    \label{fig:pi_model}
\end{figure}

This subsection focuses on electrical lines, although the same modelling principles can be extended to other branches such as transformers, provided that suitable branch admittance models are used. Each electrical line between nodes \(i\) and \(j\) is represented by a three-phase \(\pi\)-equivalent circuit (Figure \ref{fig:pi_model}), which captures both the self-impedance of each phase and the mutual coupling between phases through the series impedance matrix \(\underline{\mathbf{Z}}_{ij}\) and the line-to-ground capacitive effects via the shunt admittance matrix \(\underline{\mathbf{Y}}_{\text{sh},ij}\).

From the series and shunt components, the branch admittance matrix \(\underline{\mathbf{Y}}_{ij}^{\text{br}}\) is defined as:
\begin{equation}
\underline{\mathbf{Y}}_{ij}^{\text{br}} =
\begin{bmatrix}
\underline{\mathbf{Z}}_{ij}^{-1} + \frac{1}{2} \underline{\mathbf{Y}}_{\text{sh},ij} & -\underline{\mathbf{Z}}_{ij}^{-1} \\
-\underline{\mathbf{Z}}_{ij}^{-1} & \underline{\mathbf{Z}}_{ij}^{-1} + \frac{1}{2} \underline{\mathbf{Y}}_{\text{sh},ij}
\end{bmatrix}
\end{equation}

This \(6 \times 6\) complex matrix relates the three-phase voltages and injected currents at both terminals of the branch. The nodal current injections at both ends of the line are given by:
\begin{equation}
\begin{bmatrix}
\underline{\mathbf{I}}_{ij} \\
\underline{\mathbf{I}}_{ji}
\end{bmatrix}
=
\underline{\mathbf{Y}}_{ij}^{\text{br}} \cdot
\begin{bmatrix}
\underline{\mathbf{V}}_i \\
\underline{\mathbf{V}}_j
\end{bmatrix}
\end{equation}

Explicitly, this expression can be expanded as:
\begin{equation}
\begin{bmatrix}
\underline{\mathbf{I}}_{ij} \\
\underline{\mathbf{I}}_{ji}
\end{bmatrix}
=
\begin{bmatrix}
\underline{\mathbf{Y}}_{ij}^{ii} & \underline{\mathbf{Y}}_{ij}^{ij} \\
\underline{\mathbf{Y}}_{ij}^{ji} & \underline{\mathbf{Y}}_{ij}^{jj}
\end{bmatrix}
\cdot
\begin{bmatrix}
\underline{\mathbf{V}}_i \\
\underline{\mathbf{V}}_j
\end{bmatrix}
\end{equation}

In this expression, the submatrices of the branch admittance matrix have the following interpretations: the diagonal blocks \(\underline{\mathbf{Y}}_{ij}^{ii}\) and \(\underline{\mathbf{Y}}_{ij}^{jj}\) represent the self-admittance matrices at nodes \(i\) and \(j\), respectively, and are given by \(\underline{\mathbf{Y}}_{ij}^{ii} = \underline{\mathbf{Y}}_{ij}^{jj} = \underline{\mathbf{Z}}_{ij}^{-1} + \frac{1}{2} \underline{\mathbf{Y}}_{\text{sh},ij}\). These include both the series impedance and half of the shunt admittance. The off-diagonal blocks \(\underline{\mathbf{Y}}_{ij}^{ij}\) and \(\underline{\mathbf{Y}}_{ij}^{ji}\) correspond to the mutual admittances between the terminals of the line and are equal to \(-\underline{\mathbf{Z}}_{ij}^{-1}\), capturing the coupling introduced by the series impedance between nodes \(i\) and \(j\).

This formulation captures all mutual interactions among the three phases and both terminals of the branch. When the branch corresponds to a transformer, the expressions for \(\underline{\mathbf{Y}}_{ij}^{\text{br}}\) must be adapted according to the winding configuration (e.g., wye-wye, delta-wye). Guidelines for constructing the appropriate admittance matrices for transformer models are detailed in~\cite{Zimmerman}.

\subsection{Global Nodal Admittance Matrix Construction}

The global nodal admittance matrix \(\underline{\mathbf{Y}}\) is constructed by assembling the contributions of all branch admittance matrices \(\underline{\mathbf{Y}}_{ij}^{\text{br}}\) associated with the set of electrical lines \(\mathcal{L}\). In addition to distribution lines, other branches such as distribution transformers may also contribute to \(\underline{\mathbf{Y}}\), depending on their electrical connection and model. Each local admittance matrix is placed into the global matrix based on the topological configuration of the network and the phase availability at each node. 

This assembly yields a sparse, complex-valued admittance matrix that captures the electrical coupling between all nodes and phases in the system. The resulting relation between the complex nodal current injections \(\underline{\mathbf{I}}\) and the complex nodal voltages \(\underline{\mathbf{V}}\) is given by:
\begin{equation}
\underline{\mathbf{I}} = \underline{\mathbf{Y}} \cdot \underline{\mathbf{V}}
\end{equation}

\subsection{Nodal Power-Balance Equations}

At each node \( i \) and for each phase \( \phi \), the injected complex current \(\underline{I}_i^\phi\) is computed using the global nodal admittance matrix \(\underline{\mathbf{Y}}\) and the complex voltage vector \(\underline{\mathbf{V}}\). This current represents the net flow of current injected into the network at node \(i\) and phase \(\phi\), and is given by:
\begin{equation}
\underline{I}_i^{\phi} = \sum_{j=1}^{n} \underline{\mathbf{Y}}_{ij} \cdot \underline{\mathbf{V}}_j \big|_{\{\phi\}}
\label{eq:9}
\end{equation} 
where \(\underline{\mathbf{Y}}_{ij}\) is the \((i,j)\)-th submatrix of the global admittance matrix \(\underline{\mathbf{Y}}\), with dimensions \(3 \times 3\), which captures the couplings between the three phases at nodes \(i\) and \(j\). The expression \(\underline{\mathbf{Y}}_{ij} \cdot \underline{\mathbf{V}}_j\) yields a 3-phase complex current vector, and the subscript \(\{\phi\}\) indicates the extraction of the component corresponding to phase \(\phi\).

Once the complex current \(\underline{I}_i^\phi\) is computed, the corresponding complex power injection at the node is defined as:
\begin{equation}
\underline{S}_i^{\phi} = \underline{V}_i^{\phi} \cdot (\underline{I}_i^{\phi})^*
\end{equation}

This power injection can be decomposed into its active and reactive components as follows:
\begin{align}
P_i^{\phi} &= \Re\{\underline{S}_i^{\phi}\} = P_{G,i}^{\text{conv},\phi} + P_{G,i}^{\text{pv},\phi} - P_{D,i}^{\phi} \\
Q_i^{\phi} &= \Im\{\underline{S}_i^{\phi}\} = Q_{G,i}^{\text{conv},\phi} + Q_{G,i}^{\text{pv},\phi} - Q_{D,i}^{\phi}
\end{align}

These equations ensure that, for every node and phase, the total injected power (from conventional and photovoltaic generation) equals the consumed power by the loads plus network losses. The nodal voltages \(\underline{V}_i^\phi\) are decision variables of the optimization problem, while \(\underline{I}_i^\phi\), \(\underline{S}_i^\phi\), \(P_i^\phi\), and \(Q_i^\phi\) are linear expressions expressed in terms of the decision variables and used to impose the physical constraints of the system.

\subsection{Definition of Line Power Flows}

The complex power flow from node \(i\) to node \(j\) on phase \(\phi\) is defined as:
\begin{equation}
\underline{S}_{ij}^{\phi} = \underline{V}_i^{\phi} \cdot\left( \underline{I}_{ij}^{\phi} \right)^*
\label{eq:13}
\end{equation} 

This quantity is not a decision variable but rather a linear expression that depends on the sending-end voltage and the current injected into the line \((i,j)\). Its real and imaginary components correspond to the active and reactive power flows:
\begin{align}
P_{ij}^{\phi} &= \Re\left\{ \underline{S}_{ij}^{\phi} \right\} \\
Q_{ij}^{\phi} &= \Im\left\{ \underline{S}_{ij}^{\phi} \right\}
\end{align}

The complex current injection \(\underline{I}_{ij}^{\phi}\) from node \(i\) to node \(j\) on phase \(\phi\) is not a decision variable either. It is analytically computed from the nodal voltages \(\underline{\mathbf{V}}_i\) and \(\underline{\mathbf{V}}_j\) using the branch admittance submatrices \(\underline{\mathbf{Y}}_{ij}^{ii}\) and \(\underline{\mathbf{Y}}_{ij}^{ij}\). In particular, the phase-\(\phi\) component of the current vector \(\underline{\mathbf{I}}_{ij}\) is given by:
\begin{equation}
\underline{I}_{ij}^{\phi} = \left( \underline{\mathbf{Y}}_{ij}^{ii} \cdot \underline{\mathbf{V}}_i + \underline{\mathbf{Y}}_{ij}^{ij} \cdot \underline{\mathbf{V}}_j \right)_{\{\phi\}}
\label{eq:16}
\end{equation} 

Here, subscript \(\{\phi\}\) denotes the \(\phi\)-th element of the resulting vector, corresponding to the phase of interest. The matrices \(\underline{\mathbf{Y}}_{ij}^{ii}\) and \(\underline{\mathbf{Y}}_{ij}^{ij}\) are the self and mutual admittance blocks of the full branch admittance matrix \(\underline{\mathbf{Y}}_{ij}^{\text{br}}\), as previously defined in the branch model. This formulation ensures that all quantities related to power flows are consistently derived from the decision variables, namely the complex voltages at each node and phase.

\subsection{Optimization Problem}

The Unbalanced Optimal Power Flow problem is formulated as a non-linear optimization problem whose objective is to minimize the total generation cost of conventional units. The formulation accounts for the unbalanced nature of LV distribution networks, explicitly modelling each phase independently using complex variables. 

The decision variables of the problem are the complex nodal voltages \( \underline{\mathbf{V}} \), which comprise all phase voltages \( \underline{V}_i^{\phi} \), and the power injections \( P_{G,i}^{\text{conv},\phi}, Q_{G,i}^{\text{conv},\phi}, P_{G,i}^{\text{pv},\phi}, Q_{G,i}^{\text{pv},\phi} \). The current injections \( \underline{I}_i^\phi \) and line power flows \( \underline{S}_{ij}^{\phi} \) are linear expression that can be obtained from equations~\eqref{eq:9}, \eqref{eq:13} and \eqref{eq:16}.

\begin{small}
\begin{align}
\min \quad & \sum_{i \in \mathcal{G}_c} \sum_{\phi \in \{a,b,c\}} c_i^{\text{conv}} P_{G,i}^{\text{conv},\phi}
\label{eq:ubopf_obj} \\
\text{s.t.} \quad
& P_{G,i}^{\text{conv},\phi} + P_{G,i}^{\text{pv},\phi} - P_{D,i}^{\phi}
= \Re\left\{ \underline{V}_i^{\phi} \cdot \left( \underline{I}_i^{\phi} \right)^* \right\}, 
\quad \forall i \in \mathcal{N},\ \forall \phi
\label{eq:ubopf_pbal} \\
& Q_{G,i}^{\text{conv},\phi} + Q_{G,i}^{\text{pv},\phi} - Q_{D,i}^{\phi}
= \Im\left\{ \underline{V}_i^{\phi} \cdot \left( \underline{I}_i^{\phi} \right)^* \right\}, 
\quad \forall i \in \mathcal{N},\ \forall \phi
\label{eq:ubopf_qbal} \\
& V_{\min}^{\phi} \leq |\underline{V}_i^{\phi}| \leq V_{\max}^{\phi},
\quad \forall i \in \mathcal{N},\ \forall \phi
\label{eq:ubopf_vlimits} \\
& P_{G,i}^{\text{conv},\phi,\min} \leq P_{G,i}^{\text{conv},\phi} \leq P_{G,i}^{\text{conv},\phi,\max},
\quad \forall i \in \mathcal{G}_c,\ \forall \phi
\label{eq:ubopf_pg_limits} \\
& Q_{G,i}^{\text{conv},\phi,\min} \leq Q_{G,i}^{\text{conv},\phi} \leq Q_{G,i}^{\text{conv},\phi,\max},
\quad \forall i \in \mathcal{G}_c,\ \forall \phi
\label{eq:ubopf_qg_limits} \\
& \left(P_{G,i}^{\text{pv},\phi}\right)^2 + \left(Q_{G,i}^{\text{pv},\phi}\right)^2 
\leq \left(S_{G,i}^{\text{pv},\phi}\right)^2,
\quad \forall i \in \mathcal{G}_{\text{pv}},\ \forall \phi
\label{eq:ubopf_pq_circle} \\
& -{P}_{ij,max}^{\phi} 
\leq \Re\left\{ \underline{S}_{ij}^{\phi} \right\} 
\leq {P}_{ij,max}^{\phi},
\quad \forall (i,j) \in \mathcal{L},\ \forall \phi
\label{eq:ubopf_s_limits}
\end{align}
\end{small}

The objective \eqref{eq:ubopf_obj} seeks cost minimization, which implicitly promotes maximum utilization of PV generation due to its zero marginal cost. Power balances at each node and phase are enforced by \eqref{eq:ubopf_pbal}–\eqref{eq:ubopf_qbal}.  The voltage magnitude limits are enforced by~\eqref{eq:ubopf_vlimits}, while the operational bounds for conventional generators are given by~\eqref{eq:ubopf_pg_limits} and~\eqref{eq:ubopf_qg_limits}. Constraint \eqref{eq:ubopf_pq_circle} bounds the PV inverter outputs within their apparent power limit \( S_{G,i}^{\text{pv},\phi} \), which varies with solar irradiance. Finally, \eqref{eq:ubopf_s_limits} imposes thermal constraints on the active power flows in each line and phase.

\subsection{Modelling Photovoltaic Generation in UBOPF}

The proposed UBOPF formulation explicitly incorporates PV generation by introducing independent decision variables for both active and reactive power injections from inverters at each node and phase of the network. This approach is general and suitable for scenarios where the inverters operate with full control capabilities, enabling the adjustment of their operating point within the $P$-$Q$ plane defined by the inverter’s maximum apparent power limit. In this framework, constraint~\eqref{eq:ubopf_pq_circle} ensures that the complex power injection from each PV inverter does not exceed its available apparent power capacity determined by the instantaneous solar irradiance level.

In scenarios where a dynamic active power limitation strategy is enforced—commonly used to mitigate voltage rise problems—the inverters are assumed to operate at unity power factor. Under this operating mode, the reactive power injection from PV units must be zero across all nodes and phases. This behaviour is modelled by adding equality constraints of the form:
\begin{equation}
Q_{G,i}^{\text{pv},\phi} = 0 \quad \forall i \in \mathcal{G}_{\text{pv}}, \forall \phi 
\end{equation}
which restrict the operation to a purely active power injection regime.

As a result, constraint in~\eqref{eq:ubopf_pq_circle} reduces to a one-dimensional condition that simply limits the magnitude of the active power injection:
\begin{equation}
P_{G,i}^{\text{pv},\phi} \leq S_{G,i}^{\text{pv},\phi}
\end{equation}

This simplification not only reduces the dimensionality of the feasible region but also reflects practical limitations of standard residential inverters, which are often configured to operate without reactive power support. The UBOPF formulation is thus flexible enough to accommodate both advanced inverter control modes and conventional unity power factor operation, allowing a detailed evaluation of their impact on network performance and PV hosting capacity.

\subsection{Implementation in Julia}

The UBOPF formulation described above has been implemented as a computational program developed in the \texttt{Julia} programming language. The optimization problem is modelled using the \texttt{JuMP} package, which provides a high-level algebraic interface for mathematical programming, and solved using the \texttt{Ipopt} solver, which is designed for large-scale non-linear optimization.

Due to the use of full AC power flow equations and the explicit modelling of unbalanced three-phase systems, the resulting optimization problem is both non-linear and non-convex. These characteristics arise from the bilinear terms in the power balance constraints and the modulus expressions involved in voltage and apparent power limits. As a result, solving the UBOPF requires advanced numerical techniques and careful handling of solver settings to ensure convergence to feasible and meaningful solutions.


Despite the computational challenges posed by the problem’s complexity, the use of modern tools like \texttt{JuMP} and \texttt{Ipopt} allows for efficient and scalable solution of realistic distribution networks. This implementation serves as a foundation for the case studies and analysis presented in the subsequent sections.


\section{Case Studies} \label{sec:case_studies}
This section presents a comprehensive set of case studies evaluating the effectiveness of inverter control strategies, aiming to mitigate overvoltages and maximize PV integration in residential low-voltage distribution networks.

The analysis is structured in three parts: (1) a detailed description of the test feeder and its operational conditions; (2) evaluation of the inverter control approaches in a well-defined baseline scenario characterized by high solar irradiance and low residential demand; and (3) analysis of critical operational factors through the definition of supplementary scenarios which assess how key variables affect voltage regulation and control strategy performance.

The case studies presented herein have been thoroughly simulated using the UBOPF program implemented in \texttt{Julia}, covering all scenarios and control strategies described above. To ensure transparency and facilitate validation as well as further research, the complete simulation code, input data and relevant results are publicly available in a dedicated GitHub repository \cite{github_repo}.

\subsection{Test Network Modelling}
\begin{figure}[h]
    \centering
    \includegraphics[width=1\linewidth]{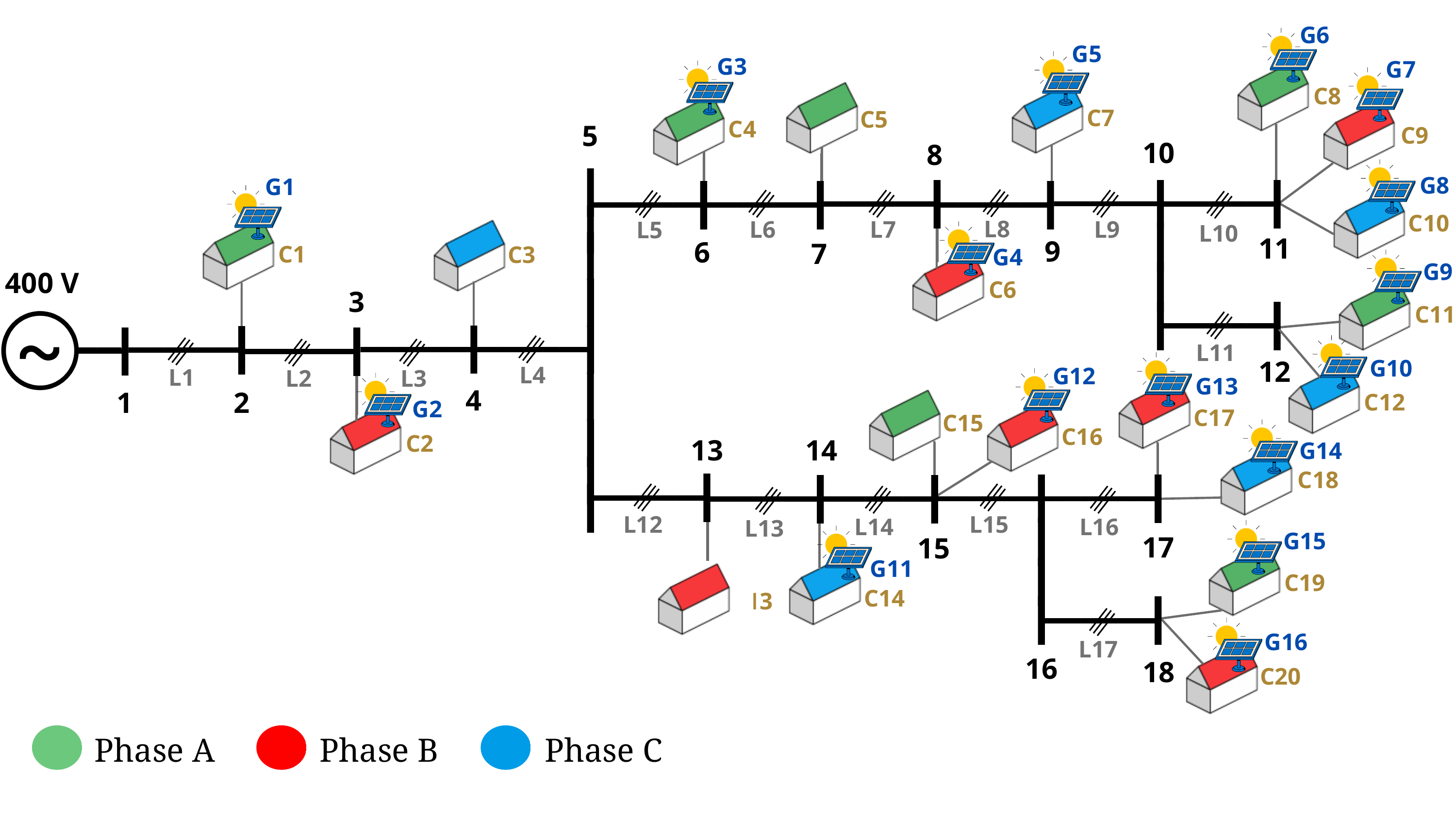}
    \caption{Reference low-voltage distribution network with 18 nodes, 20 residential loads (C1–C20) and 16 PV inverters (G1–G16).}
    \label{fig:network}
\end{figure}

The reference system is a 400~V three-phase radial distribution feeder consisting of 18 nodes and 17 identical line segments. It supplies 20 single-phase residential loads connected in an unbalanced manner across the three phases. To emulate a high PV penetration scenario, distributed generation is assigned to 80\% of the loads, resulting in 16 single-phase PV inverters unevenly connected across the network. The network topology is shown in Figure~\ref{fig:network}.

All lines are modelled using full $\pi$-equivalent three-phase representations including mutual coupling. Electrical parameters are based on typical underground cables commonly used in European LV networks, as reported in~\cite{Churkin}. The upstream system is represented as a balanced three-phase ideal voltage source connected at node~1.

The 20 residential loads are represented as constant PQ demand and allocated across the three phases to emulate typical unbalanced conditions found in low-voltage networks. Load consumption values are derived from a measurement-based dataset of residential demand profiles~\cite{Rodriguez}.

\subsection{Baseline Scenario Analysis}

\begin{figure}[h]
    \centering
    \includegraphics[width=1\linewidth]{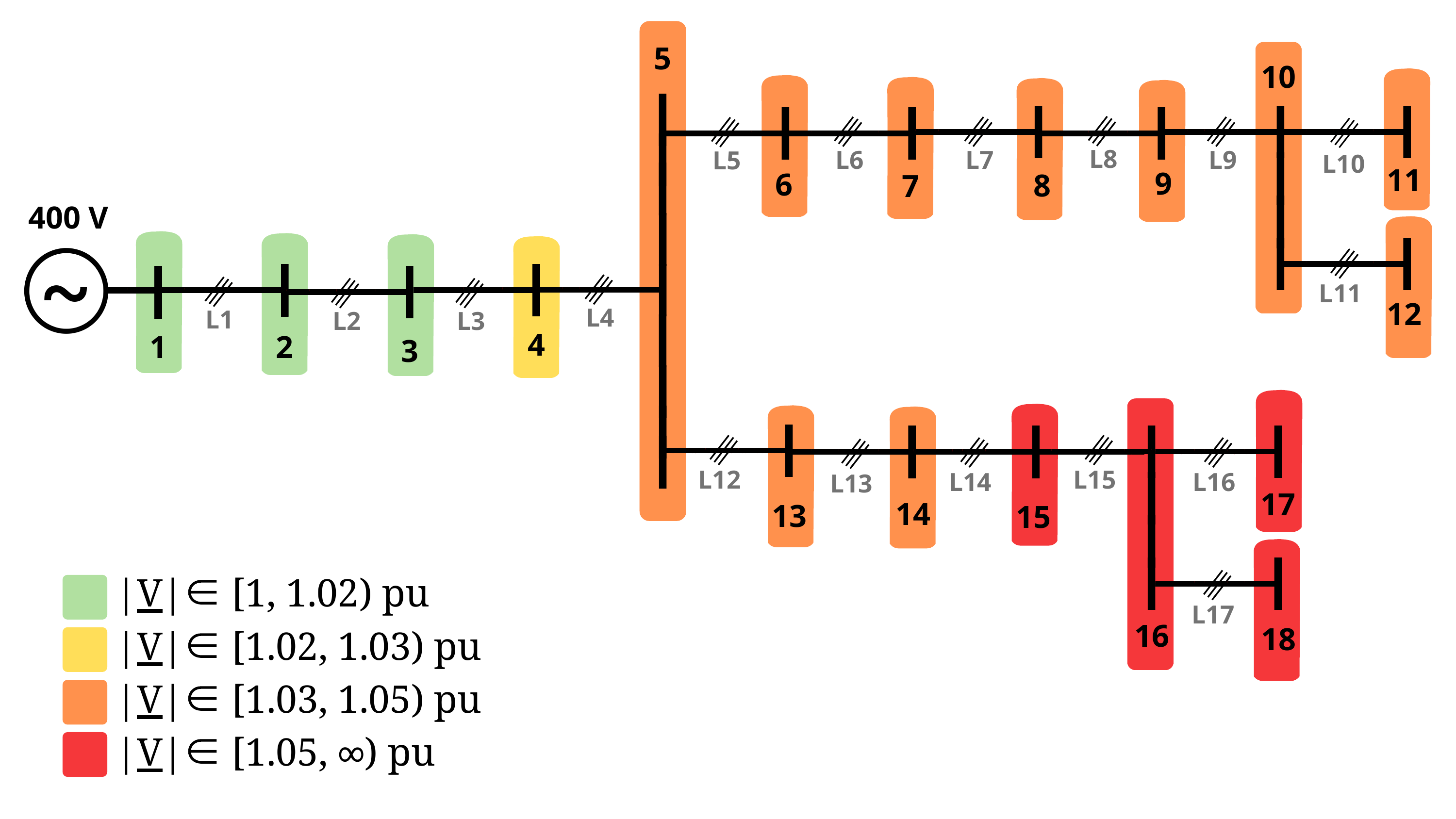}
    \caption{Voltage profile in phase B without control. Overvoltages occur at nodes with high PV injection and weak local load.}
    \label{fig:volt_profile_noctrl}
\end{figure}

The baseline scenario corresponds to a critical operating condition characterized by high solar irradiance and low residential demand, under which all PV inverters operate at their maximum available power. Without any voltage control mechanism, this condition leads to overvoltages at several nodes—particularly those connected to phase B—due to the excess active power injected by PV units and insufficient local consumption. The resulting voltage rise along the feeder causes multiple nodes to exceed the standard limit of 1.03~pu, as illustrated in Figure~\ref{fig:volt_profile_noctrl}. 

To mitigate this phenomenon and maintain compliance with the regulatory voltage limits—previously set to $\pm 3\%$ (0.97–1.03~pu), as commonly adopted by distribution system operators—two alternative control approaches are tested on the same operating scenario using the proposed UBOPF formulation. These approaches consist of: (i) dynamic curtailment of active power only, and (ii) coordinated control of both active and reactive power in PV inverters.

In this scenario, all PV inverters are assigned the same apparent power rating of 2500~VA, a typical value for single-phase rooftop systems~\cite{Navarro}. The impact of each method on system performance is summarized in Table~\ref{tab:comparison_strategies}, comparing total PV injection, curtailed energy and the maximum nodal voltage.

\begin{table}[h]
\centering
\caption{Performance of the different control approaches under the baseline scenario}
\label{tab:comparison_strategies}
\resizebox{\columnwidth}{!}{%
\begin{tabular}{lccc}
\toprule
\textbf{Control approach} & $P_{\text{PV,total}}$ (kW) & Curtailment (kW) & $|V|_{\text{max}}$ (pu) \\
\midrule
No control                   & 40.00  & 0.0     & \cellcolor{red!20}1.056 \\
P-only control    & 35.99  & 4.01 (11\%) & 1.030 \\
P and Q control & 37.56  & 2.44 (6.5\%)  & 1.030 \\
\bottomrule
\end{tabular}%
}
\end{table}

In the absence of control, no curtailment occurs and all photovoltaic generators inject their full available power as active power (16 units × 2500~W = 40~kW). Both control approaches effectively cancel overvoltages and ensure compliance with voltage limits (1.03 pu). However, the coordinated control of $P$ and $Q$ proves to be more efficient in terms of renewable energy utilization, reducing curtailment by nearly 40\% compared to active power curtailment alone.

A more detailed insight can be obtained by analyzing the operation of specific PV inverters under both control strategies. Table~\ref{tab:injections_comparison} presents the active ($P$) and reactive ($Q$) power injections of selected units. Under active power curtailment, all PV inverters are forced to reduce their output proportionally to maintain voltages within the permissible range. No reactive power is injected or absorbed, which limits the flexibility of this approach.

\begin{table}[h]
\centering
\caption{Power injection by selected PV inverters under different control approaches}
\label{tab:injections_comparison}
\resizebox{\columnwidth}{!}{%
\begin{tabular}{cccccc}
\toprule
\multirow{2}{*}{\textbf{PV Unit}} & \multicolumn{2}{c}{P-only control} & \multicolumn{2}{c}{P and Q control} & $\Delta P_\text{PV}$ (W) \\
\cmidrule(lr){2-3} \cmidrule(lr){4-5}
 & $P$ (W) & $Q$ (var) & $P$ (W) & $Q$ (var) & \\
\midrule
G1  & 2500.0 & 0.0   & 2498.6 & $-84.2$  & $-1.4$ \\
G5  & 2500.0 & 0.0   & 2473.2 & $-302.1$ & $-26.8$ \\
G9  & 2315.8 & 0.0   & 2397.9 & $-293.6$ & $+82.1$ \\
G10 & 2297.8 & 0.0   & 2368.7 & $-312.3$ & $+70.9$ \\
\bottomrule
\end{tabular}%
}
\end{table}

However, when both active and reactive power are coordinated, the UBOPF optimizer adjusts the $Q$ injections of each inverter based on local network conditions. This allows for a more tailored and efficient operation: units at voltage-sensitive locations may absorb reactive power to reduce local voltage and thereby increase—or at least maintain—their $P$ injection.

For example, inverter G9 increases its active power output from 2315.8~W to 2397.9~W while absorbing 293.6~var, and G10 follows a similar pattern. This behaviour highlights how reactive support can relax local voltage constraints and enhance overall generation.

Conversely, inverters like G1 and G5—which under active power curtailment were injecting their full 2500~W capacity—must reduce their active output slightly under the coordinated strategy (e.g., to 2473.2~W for G5) to enable substantial reactive power injection. This scenario illustrates a key equity issue: some prosumers may incur a reduction in usable energy to support system-wide voltage regulation. 

Therefore, this type of operation should be recognized as a grid-supporting ancillary service, and regulatory frameworks should evolve to provide appropriate remuneration mechanisms that incentivize participation and ensure fairness. Coordinated control of active and reactive power not only guarantees voltage compliance but also promotes a more flexible and efficient use of distributed photovoltaic generation. It enhances the network’s hosting capacity while minimizing the need for curtailment, particularly under unbalanced or low-demand operating conditions.

In the baseline scenario analysed, active power curtailment remains moderate (below 5\%), confirming that effective voltage regulation can be achieved without significantly limiting solar output. Nevertheless, one may wonder whether higher curtailment levels could discourage participation or even lead to the abandonment of PV installations. In practice, this outcome is unlikely. Even in curtailment situations, PV systems continue to inject a considerable share of their available power into the grid, thereby maintaining economic profitability. Moreover, the rapid deployment of distributed battery storage systems provides an increasingly viable pathway for utilizing curtailed energy—either by storing it locally or exporting it later when network conditions allow. Thus, rather than promoting PV abandonment, the proposed UBOPF-based coordination strategy fosters a more reliable, economically sustainable, and future-ready integration of distributed solar generation.

\subsection{Additional Scenarios for the Analysis of Critical Factors}

To deepen the understanding of the mechanisms leading to overvoltages in LV networks with high PV penetration and to assess the robustness of inverter-based control strategies, a series of additional scenarios has been defined. These scenarios (S1 to S5) are systematic variations of the baseline configuration. Each scenario alters a single system parameter, enabling an isolated analysis of its impact on voltage profiles and control effectiveness.

The analysis targets three key aspects: (i) the balance between local generation and demand, (ii) the reactive power behavior of residential loads, and (iii) the phase-level distribution of consumption. By decoupling these influences, the study provides a granular view of critical conditions that exacerbate or mitigate overvoltage phenomena.

\subsubsection{Impact of the Generation–Demand Balance}

\begin{figure}[h!]
\centering
\begin{tikzpicture}
\begin{axis}[
    ybar,
    bar width=15pt,
    width=0.45\textwidth,
    height=0.3\textwidth,
    enlarge x limits=0.3,
    ymin=0, ymax=55,
    ylabel={Curtailment (\%)},
    symbolic x coords={Baseline, S1, S2},
    xtick=data,
    xtick style={draw=none},
    ytick style={draw=none},
    axis x line*=bottom,
    axis y line*=left,
    grid=major,
    grid style={dashed, gray!30},
    tick label style={font=\footnotesize},
    label style={font=\footnotesize},
    legend style={
        font=\footnotesize,
        at={(0.5,1.05)}, anchor=south,
        legend columns=2,
        column sep=10pt,
        draw=none,
        fill=none
    },
    legend image code/.code={
        \draw[fill=#1, draw=none] (0cm,-0.1cm) rectangle (0.3cm,0.1cm);
    }
]
\definecolor{pastelblue}{RGB}{140,170,230}
\definecolor{pastelgreen}{RGB}{160,210,160}

\addplot+[
    ybar, 
    fill=pastelblue, 
    draw=none,
    nodes near coords,
    nodes near coords style={font=\footnotesize, color=black}
] coordinates {(Baseline,10) (S1,47.8) (S2,2.9)};
\addlegendentry{P-only control}

\addplot+[
    ybar, 
    fill=pastelgreen, 
    draw=none,
    nodes near coords,
    nodes near coords style={font=\footnotesize, color=black}
] coordinates {(Baseline,6.1) (S1,35.9) (S2,1.2)};
\addlegendentry{Coordinated P+Q control}

\end{axis}
\end{tikzpicture}
\caption{\footnotesize Curtailment percentage under P-only and coordinated P+Q control strategies in the baseline scenario and scenarios S1 (`Excess PV Generation + Low Demand') and S2 (`High Residential Demand').}
\label{fig:curtailment_comparison}
\end{figure}
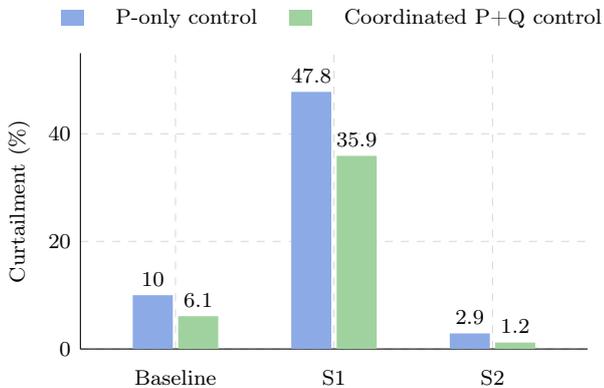

Table~\ref{tab:bar_pv_demand} compares the relative levels of PV generation and residential demand in the baseline scenario and two complementary cases. The baseline already represents a moderately critical situation, with high solar irradiance and relatively low load levels. Scenario~S1 increases the severity of this mismatch by amplifying PV capacity and reducing demand, whereas Scenario~S2 considers a higher consumption pattern that offsets PV surplus and alleviates overvoltage risk. Simulation results are included in Table~\ref{tab:results_S1_S2} and Figure~\ref{fig:curtailment_comparison}.

\begin{table}[h!]
\centering
\caption{\footnotesize Comparison of operating conditions in baseline scenario and scenarios S1 (`Excess PV Generation + Low Demand') and S2 (`High Residential Demand').}
\label{tab:bar_pv_demand}
\resizebox{0.95\columnwidth}{!}{
\begin{tabular}{lccc}
\toprule
\textbf{Metric} & \textbf{Baseline} & \textbf{S1} & \textbf{S2} \\
\midrule
PV generation (\%)                & 100\%     & 160\%     & 100\%     \\
Residential demand (\%)          & 100\%     & 50\%      & 150\%     \\
Available PV capacity (kVA)       & 40    & 64     & 40    \\
\bottomrule
\end{tabular}
}
\end{table}

\medskip

\noindent\textbf{S1 – `Excess PV Generation + Low Demand'.}  
This scenario reflects critical daytime periods when solar irradiance is high but residential demand is low—typically on weekdays during spring or autumn. To emulate this situation, the rated apparent power of each PV system was increased by 60\% (from 2500~VA to 4000~VA), while active and reactive loads were reduced by 50\%. As a result, the system operates under a net export condition. Simulation results reveal a sharp increase in voltage magnitudes, with phase~B reaching 1.114~pu in the absence of control. Substantial curtailment is required to maintain voltage within limits: 30.6~kW of PV generation (47.8\% of the available power) must be curtailed under pure active power control. In contrast, coordinated $P+Q$ control reduces this value to 22.96~kW (35.9\%), enabling greater solar utilization despite the severity of voltage stress. These findings underscore the importance of coordinated control strategies in mitigating voltage issues. Nevertheless, they also emphasize the need for complementary solutions, such as demand response, energy storage, or grid reinforcement, when facing widespread PV deployment under light-load conditions.

\medskip

\noindent\textbf{Scenario S2 – `High Residential Demand'.}  
This case explores how growing trends in household electrification—such as heat pumps, electric vehicles and induction cooking—can contribute positively to voltage stability. In this scenario, loads are increased by 50\% compared to the baseline, while PV generation remains unchanged. The network becomes more absorptive, leading to a voltage decrease (1.043~pu maximum without control). Curtailment needs are minimal in this scenario: only 2.9\% under $P$-only control and just 1.2\% with $P+Q$ control. These results demonstrate that increased local demand can act as a natural buffer against overvoltages and facilitate PV  integration with minimal energy loss.

\begin{table}[h!]
\centering
\caption{\footnotesize Comparison of key results between the baseline scenario and scenarios S1 (`Excess PV Generation + Low Demand') and S2 (`High Residential Demand').}
\label{tab:results_S1_S2}
\resizebox{0.9\columnwidth}{!}{
\begin{tabular}{lccc}
\toprule
& \textbf{Baseline} & \textbf{S1} & \textbf{S2} \\
\midrule
\multicolumn{4}{l}{\textbf{Without control}} \\
$P_{\text{PV,total}}$ injected (kW)      & 40     & 64     & 40     \\
$|\underline{V}_{\text{max}}|$ (V)   & 243.99    & 257.33    & 241.01    \\
$|\underline{V}_{\text{max}}|$ (pu)   & 1.056     & 1.114     & 1.043     \\
\midrule
\multicolumn{4}{l}{\textbf{With P-only control}} \\
$P_{\text{PV,total}}$ injected k(W)   & 35.99   & 33.40   & 38.84   \\
Curtailment (kW)                      & 4.01    & 30.60   & 1.16    \\
\midrule
\multicolumn{4}{l}{\textbf{With coordinated P+Q control}} \\
$P_{\text{PV,total}}$ injected (kW)   & 37.56  & 41.04   & 39.52   \\
Curtailment (kW)                      & 2.44    & 22.96   & 0.48     \\
\bottomrule
\end{tabular}
}
\end{table}

\medskip
\noindent Overall, the results in Table~\ref{tab:results_S1_S2} and Figure~\ref{fig:curtailment_comparison} demonstrate that coordinated $P+Q$ reduces PV curtailment more effectively than active power limitation alone, especially under high-export conditions with large generation–demand mismatches. These results highlight the role of reactive power capability in inverter-based DER and the importance of electrification in enhancing voltage management and hosting capacity.


\subsubsection{Impact of Capacitive Behaviour of Residential Loads}

The increasing presence of power electronics in residential environments is fundamentally reshaping the reactive power characteristics of household consumption. While low-voltage networks have historically been dominated by inductive loads—such as refrigerators, washing machines, and air conditioning motors—the current trend is shifting toward more capacitive behaviour due to the proliferation of electronic devices.

\begin{figure}[h!]
    \centering
    \includegraphics[width=0.99\linewidth]{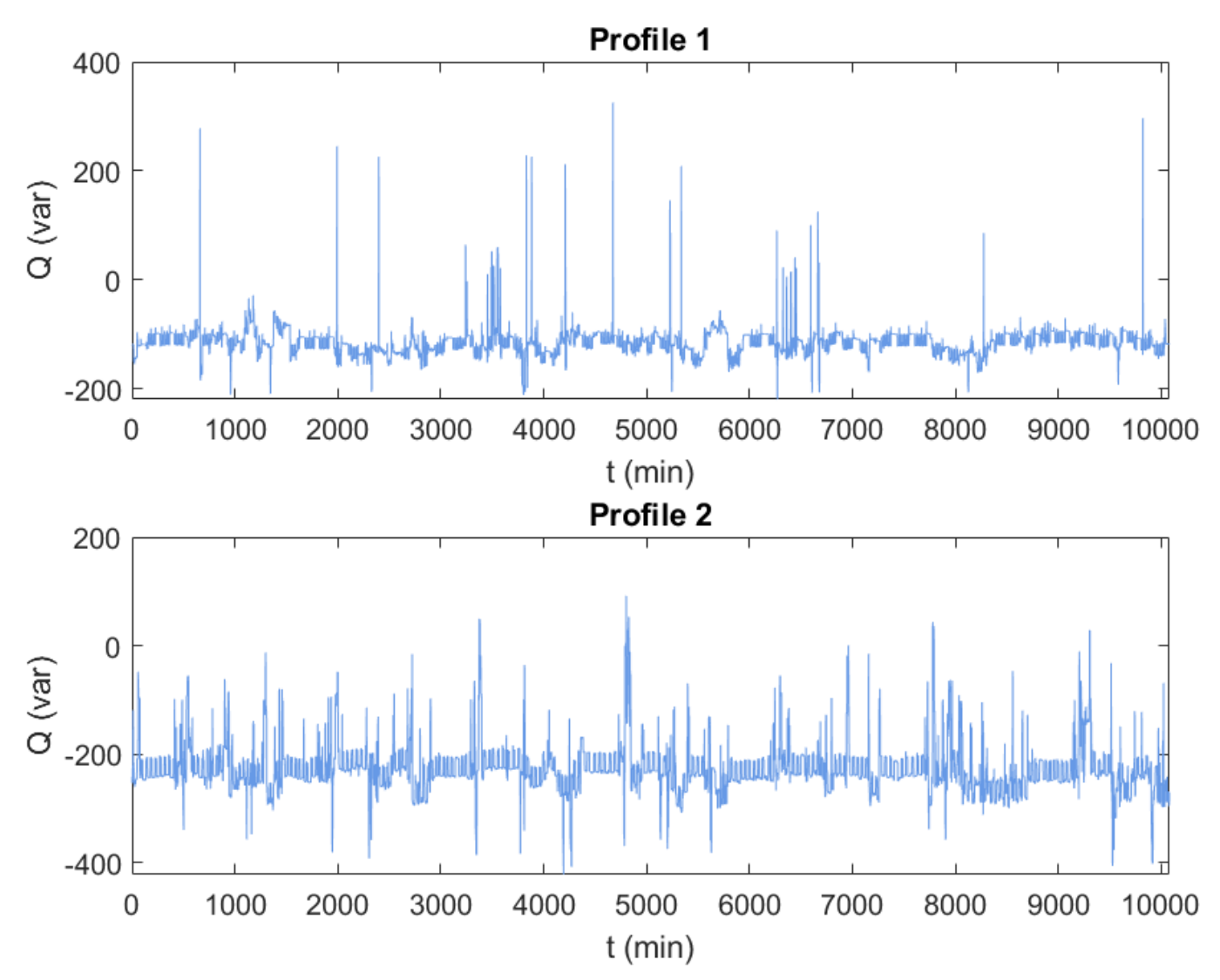}
    \caption{\footnotesize Reactive power profiles measured in two residential homes over a typical week, illustrating frequent capacitive behaviour \cite{Rodriguez}.}
    \label{fig:residential_q_profiles}
\end{figure}

Modern appliances, including LED lighting, TVs, computer chargers and devices with switch-mode power supplies, often exhibit near-unity or even leading power factor, especially when lightly loaded. These internal characteristics, such as capacitive input filters and high-frequency rectifiers, contribute to a net capacitive load profile at the household level. This phenomenon has already been reported in previous studies focusing on residential load modelling~\cite{Tazky} and is further corroborated by real-world measurements of reactive power in residential networks, shown in Figure~\ref{fig:residential_q_profiles} ~\cite{Rodriguez}.

This ongoing transformation has critical implications for voltage regulation and PV integration in distribution networks. To evaluate its impact, two additional scenarios were defined: `Half Capacitive' (Scenario~S3), where 50\% of the residential loads exhibit capacitive reactive power behaviour; and `Full Capacitive' (Scenario~S4), in which this capacitive character is extended to all residential loads.
The progression of this shift is illustrated in Table~\ref{tab:capacitive_behavior}, which summarizes the share of capacitive loads modelled in each case. Both scenarios are constructed as modifications of the baseline case, which assumes purely inductive consumption.

\begin{table}[h!]
\centering
\caption{\footnotesize Comparison of operating conditions in baseline scenario and scenarios S3 (`Half Capacitive') and S4 (`Full Capacitive').}
\label{tab:capacitive_behavior}
\resizebox{0.96\columnwidth}{!}{
\begin{tabular}{lccc}
\toprule
\textbf{Metric} & \textbf{Baseline} & \textbf{S3} & \textbf{S4} \\
\midrule
Capacitive behaviour (\%)                & 0\%     & 50\%     & 100\%     \\
PV generation (\%)                & 100\%     & 100\%     & 100\%     \\
Available PV capacity (kVA)       & 40    & 40     & 40   \\
\bottomrule
\end{tabular}
}
\end{table}

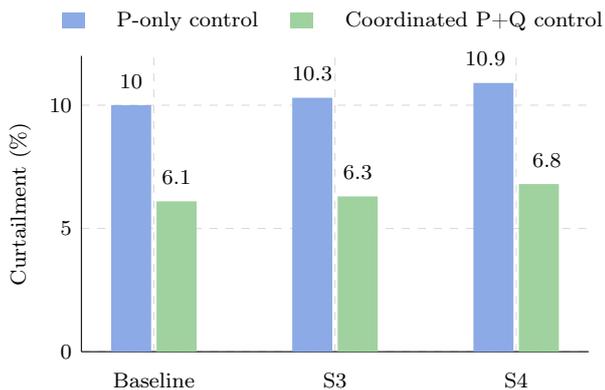
\begin{figure}[h!]
\centering
\begin{tikzpicture}
\begin{axis}[
    ybar,
    bar width=15pt,
    width=0.45\textwidth,
    height=0.3\textwidth,
    enlarge x limits=0.2,
    ymin=0, ymax=12,
    ylabel={Curtailment (\%)},
    symbolic x coords={Baseline, S3, S4},
    xtick=data,
    xtick style={draw=none},
    ytick style={draw=none},
    axis x line*=bottom,
    axis y line*=left,
    grid=major,
    grid style={dashed, gray!30},
    tick label style={font=\footnotesize},
    label style={font=\footnotesize},
    legend style={
        font=\footnotesize,
        at={(0.5,1.05)}, anchor=south,
        legend columns=2,
        column sep=10pt,
        draw=none,
        fill=none
    },
    legend image code/.code={
        \draw[fill=#1, draw=none] (0cm,-0.1cm) rectangle (0.3cm,0.1cm);
    }
]

\definecolor{pastelblue}{RGB}{140,170,230}
\definecolor{pastelgreen}{RGB}{160,210,160}

\addplot+[
    ybar, fill=pastelblue, draw=none,
    nodes near coords style={font=\footnotesize, color=black},
    nodes near coords={
        \ifnum\coordindex=2
            \node[font=\footnotesize, color=black, above, xshift=-3pt] {\pgfmathprintnumber{\pgfplotspointmeta}};
        \else
            \node[font=\footnotesize, color=black, above] {\pgfmathprintnumber{\pgfplotspointmeta}};
        \fi
    }
] coordinates {(Baseline,10) (S3,10.3) (S4,10.9)};
\addlegendentry{P-only control}

\addplot+[
    ybar, fill=pastelgreen, draw=none,
    nodes near coords style={font=\footnotesize, color=black},
    nodes near coords={
        \ifnum\coordindex=2
            \node[font=\footnotesize, color=black, above, xshift=3pt] {\pgfmathprintnumber{\pgfplotspointmeta}};
        \else
            \node[font=\footnotesize, color=black, above] {\pgfmathprintnumber{\pgfplotspointmeta}};
        \fi
    }
] coordinates {(Baseline,6.1) (S3,6.3) (S4,6.8)};
\addlegendentry{Coordinated P+Q control}

\end{axis}
\end{tikzpicture}
\caption{\footnotesize Curtailment percentage under P-only and coordinated P+Q control strategies in the baseline scenario and scenarios S3 (`Half Capacitive') and S4 (`Full Capacitive').}
\label{fig:curtailment_comparison_S3_S4}
\end{figure}

Importantly, active power demand and PV generation remain unchanged in these simulations. Only the sign of the reactive power component of the load is inverted in S3 and S4. As shown in Table~\ref{tab:results_S3_S4}, even a partial shift to capacitive demand causes a measurable voltage increase across the network, as the resulting injection of reactive power from the loads leads to a greater upward contribution in the voltage rise expression~\eqref{eq:volt_rise}. The maximum voltage rises from 1.056~pu in the baseline to 1.057~pu and 1.058~pu in scenarios S3 and S4, respectively. This subtle change translates into a clear increase in curtailed PV energy, both under P-only and coordinated \(P+Q\) control, as visualized in Figure~\ref{fig:curtailment_comparison_S3_S4}.

\begin{table}[h!]
\centering
\caption{\footnotesize Comparison of key results between the baseline scenario and scenarios S3 (`Half Capacitive') and S4 (`Full Capacitive').}
\label{tab:results_S3_S4}
\resizebox{0.9\columnwidth}{!}{
\begin{tabular}{lccc}
\toprule
& \textbf{Baseline} & \textbf{S3} & \textbf{S4} \\
\midrule
\multicolumn{4}{l}{\textbf{Without control}} \\
$P_{\text{PV,total}}$ injected (kW)      & 40     & 40     & 40 \\
$|\underline{V}_{\text{max}}|$ (V)    & 243.99   & 244.10    & 244.31    \\
$|\underline{V}_{\text{max}}|$ (pu)   & 1.056    & 1.057     & 1.058     \\
\midrule
\multicolumn{4}{l}{\textbf{With P-only control}} \\
$P_{\text{PV,total}}$ injected (kW)   & 35.99   & 35.90   & 35.63   \\
Curtailment (kW)                      & 4.01    & 4.10    & 4.37    \\
\midrule
\multicolumn{4}{l}{\textbf{With coordinated P+Q control}} \\
$P_{\text{PV,total}}$ injected (kW)   & 37.56   & 37.50   & 37.28  \\
Curtailment (kW)                      & 2.44    & 2.50    & 2.72    \\
\bottomrule
\end{tabular}
}
\end{table}

\vspace{0.5em}

Overall, these results confirm that the shift toward capacitive residential demand—driven by the widespread adoption of LED lighting and modern electronic appliances—contributes to voltage rise and undermines the network’s ability to integrate PV generation. This effect, while not as dramatic as that caused by mismatched generation and demand, is systematic and should not be overlooked. Its relevance is expected to grow with the increasing penetration of modern household appliances.

\subsubsection{Impact of Phase Imbalance in Load Distribution}

Low-voltage residential networks are predominantly composed of single-phase connections, which often leads to unequal distribution of loads across the three phases. This inherent phase imbalance can significantly affect voltage profiles and the performance of control strategies in PV-integrated systems. To investigate this effect,  a new scenario, denoted as `Phase Imbalance' (Scenario~S5), is defined by introducing a deliberate asymmetry in the allocation of residential demand.

While the baseline scenario features a reasonably balanced distribution—with 35.9\% of active power in phase~\textbf{a}, 31.0\% in phase~\textbf{b}, and 33.2\% in phase~\textbf{c}—Scenario~S5 introduces a deliberate imbalance by increasing the demand in phase~\textbf{a} to 40.9\% and reducing it in phase~\textbf{b} to 25.8\%, as shown in Table~\ref{tab:phase_distribution_S5}. The reactive power distribution remains nearly unchanged, with only minor variations among phases. As a result, the observed increase in voltage and PV curtailment is primarily due to the redistribution of active power. This outcome aligns with the typical behaviour of low-voltage networks, where the resistance-to-reactance ratio (\(R/X\)) is significantly greater than one, making voltage rise more sensitive to active than reactive power flows. Importantly, the total demand and PV generation remain unchanged between both scenarios.

\begin{table}[h!]
\centering
\caption{\footnotesize Phase-wise distribution of residential active and reactive demand in the baseline scenario and Scenario~S5 (`Phase Imbalance').}
\label{tab:phase_distribution_S5}
\resizebox{0.7\columnwidth}{!}{
\renewcommand{\arraystretch}{1.15}
\begin{tabular}{c cc cc}
\toprule
\textbf{Phase} & \multicolumn{2}{c}{\textbf{Baseline}} & \multicolumn{2}{c}{\textbf{S5}} \\
               & \%$P_D$ & \%$Q_D$ & \%$P_D$ & \%$Q_D$ \\
\midrule
a & 35.9 & 36.9 & 40.9 & 36.9 \\
b & 31.0 & 34.9 & 25.8 & 32.9 \\
c & 33.2 & 28.2 & 33.3 & 30.2 \\
\bottomrule
\end{tabular}
}
\end{table}

Despite identical generation and total demand conditions, the asymmetric load distribution introduced in Scenario~S5 leads to a localized voltage rise in the least loaded phase. As detailed in Table~\ref{tab:results_S5}, the network’s maximum voltage increases from 1.056~pu in the baseline scenario to 1.066~pu in S5. This voltage elevation directly limits the hosting capacity for PV systems, particularly under active power curtailment. In that case, curtailed energy rises from 4.01~kW to 4.90~kW, whereas with coordinated \(P+Q\) control, curtailment increases more moderately, from 2.44~kW to 2.59~kW.

\begin{table}[h!]
\centering
\caption{\footnotesize Comparison of key results between the baseline scenario and scenario~S5 (`Phase Imbalance').}
\label{tab:results_S5}
\resizebox{0.75\columnwidth}{!}{
\begin{tabular}{lcc}
\toprule
& \textbf{Baseline} & \textbf{S5} \\
\midrule
\multicolumn{3}{l}{\textbf{Without control}} \\
$P_{\text{PV,total}}$ injected (kW)      & 40     & 40 \\
$|\underline{V}_{\text{max}}|$ (V)  & 243.99    & 246.08    \\
$|\underline{V}_{\text{max}}|$ (pu) & 1.056     & 1.066     \\
\midrule
\multicolumn{3}{l}{\textbf{With P-only control}} \\
$P_{\text{PV,total}}$ injected (kW)                & 35.99   & 35.10  \\
Curtailment (kW)                                    & 4.01    & 4.90    \\
\midrule
\multicolumn{3}{l}{\textbf{With coordinated P+Q control}} \\
$P_{\text{PV,total}}$ injected (kW)                & 37.56   & 37.41   \\
Curtailment (kW)                                    & 2.44    & 2.59    \\
\bottomrule
\end{tabular}
}
\end{table}

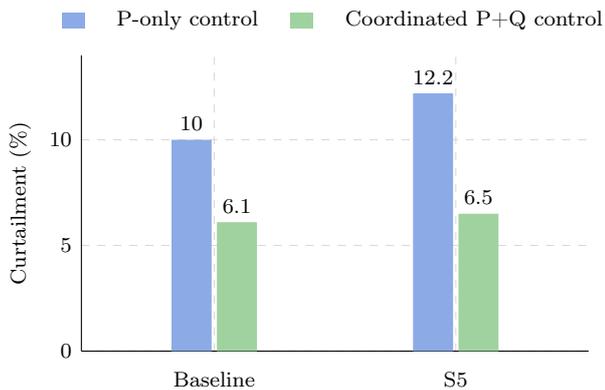
\begin{figure}[h!]
\centering
\begin{tikzpicture}
\begin{axis}[
    ybar,
    bar width=15pt,
    width=0.45\textwidth,
    height=0.3\textwidth,
    enlarge x limits=0.55,
    ymin=0, ymax=14,
    ylabel={Curtailment (\%)},
    symbolic x coords={Baseline, S5},
    xtick=data,
    xtick style={draw=none},
    ytick style={draw=none},
    axis x line*=bottom,
    axis y line*=left,
    grid=major,
    grid style={dashed, gray!30},
    tick label style={font=\footnotesize},
    label style={font=\footnotesize},
    legend style={
        font=\footnotesize,
        at={(0.5,1.05)}, anchor=south,
        legend columns=2,
        column sep=10pt,
        draw=none,
        fill=none
    },
    legend image code/.code={
        \draw[fill=#1, draw=none] (0cm,-0.1cm) rectangle (0.3cm,0.1cm);
    }
]
\definecolor{pastelblue}{RGB}{140,170,230}
\definecolor{pastelgreen}{RGB}{160,210,160}
\addplot+[
    ybar, fill=pastelblue, draw=none,
    nodes near coords style={font=\footnotesize, color=black},
    nodes near coords
] coordinates {(Baseline,10) (S5,12.2)};
\addlegendentry{P-only control}
\addplot+[
    ybar, fill=pastelgreen, draw=none,
    nodes near coords style={font=\footnotesize, color=black},
    nodes near coords
] coordinates {(Baseline,6.1) (S5,6.5)};
\addlegendentry{Coordinated P+Q control}
\end{axis}
\end{tikzpicture}
\caption{\footnotesize Curtailment percentage under P-only and coordinated $P+Q$ control strategies in the baseline scenario and Scenario~S5 (`Phase Imbalance').}
\label{fig:curtailment_comparison_S5}
\end{figure}

These results are visualized in Figure~\ref{fig:curtailment_comparison_S5}, which compares the percentage of curtailed PV energy across both strategies. Under P-only control, the phase loading in Scenario~S5 leads to a curtailment level nearly 900~W higher than in the baseline scenario. The coordinated \(P+Q\) strategy also suffers a slight degradation, with curtailment increasing by approximately 150~W. These deviations, while moderate, demonstrate how uneven phase-level demand can reduce PV injection margins by locally pushing voltages beyond operational limits.

These findings highlight the relevance of phase-level load balancing in mitigating voltage rise and maximizing PV integration. Dynamic phase-switching technologies have been proposed as a viable solution to redistribute single-phase loads in real time, improving operational conditions in residential distribution networks~\cite{Shahnia}.

\section{Conclusion and Future Work} \label{sec:conclusion}
The growing penetration of distributed photovoltaic generation in low-voltage networks challenges traditional planning and operational practices. This work has demonstrated that a detailed three-phase formulation of the optimal power flow problem—UBOPF—enables not only accurate analysis of such phenomena but also effective coordination of inverter-based control strategies.

Beyond confirming the technical effectiveness of active and reactive power control, the results reveal that local conditions—such as load imbalance, low demand or capacitive power factors—can critically undermine the potential of distributed PV, even when control capabilities are available. In this context, mitigation cannot rely solely on local inverter logic or fixed curtailment policies; it requires a system-level approach that internalizes the electrical realities of unbalanced operation and actively coordinates distributed flexibility.

While the proposed UBOPF framework provides such a system-level benchmark, its direct implementation in real distribution networks remains challenging. Achieving fully centralised coordination would require extensive communication infrastructure, continuous data exchange and robust real-time optimisation, which may raise scalability, privacy and reliability concerns. Therefore, the present study should be interpreted as an analytical tool for identifying technically optimal coordination strategies rather than a ready-to-deploy operational model. A practical path forward would involve translating these optimal results into decentralised or hierarchical control architectures capable of approximating the same behaviour under realistic operating conditions.

This insight calls for a rethinking of voltage control as not merely a constraint to satisfy, but as a service to be valued. One promising direction is the integration of economic signals through Locational Marginal Prices (LMPs) adapted to distribution-level operation. By reflecting the true local impact of voltage regulation actions, LMPs could form the basis of incentive mechanisms that reward prosumers for providing voltage support precisely where and when it is needed. Incorporating such pricing signals into the UBOPF framework opens the door to market-based coordination of inverter control, aligning technical needs with economic drivers.

Future research will focus on extending the UBOPF model to account for additional real-world complexities such as partial or uncertain network information, limited observability, and communication constraints. These aspects will be addressed through decentralised and hierarchical control schemes that can approximate the global optimum under distributed conditions. Furthermore, incorporating distribution-level LMP formulations that consider phase-specific sensitivities and unbalances will allow the design of fair, transparent and efficient compensation schemes, ultimately fostering active participation of distributed resources in voltage management and enhancing the reliability and flexibility of future low-voltage networks.

\section*{Funding}
This research was supported by the Madrid Government (Comunidad de Madrid-Spain) under the Multiannual Agreement 2023-2026 with Universidad Politécnica de Madrid, `Line A - Emerging PIs' (grant number: 24-DWGG5L-33-SMHGZ1), as well as by MICIU/AEI \linebreak[4]/10.13039/501100011033 and by ERDF/EU under Project PID2022-141609OB-I00.

\section*{Acknowledgment}
The authors would like to express sincere gratitude to Alireza Zabihi, Ph.D. candidate at Universidad Politécnica de Madrid, for his valuable support and guidance during the development and implementation phases of the computational program in \texttt{Julia}. His technical insight and willingness to share his expertise greatly contributed to the successful completion of this work.


\end{document}